\documentstyle [12pt] {article}
\title{LC OSCILLATING CIRCUIT AS THE SIMPLE CLASSICAL ANALOG
                OF THE QUANTUM ZENO EFFECT}

\author{Vladan Pankovi\'c\\
Department of Physics, Faculty of Sciences, 21000 Novi Sad,\\ Trg
Dositeja Obradovi\'ca 4. , Serbia, vpankovic@if.ns.ac.yu}

\date {}
\begin{document}
\maketitle

\vspace {0.5cm} PACS number: 03.65.Ta \vspace {0.5cm}

\begin {abstract}
In this work a simple classical analog of the quantum Zeno effect
is suggested. As it is well known, in the quantum mechanics, in
the limit of the infinite series of alternative short dynamical
evolution and measurement, an unstable quantum system will never
decay, that is called quantum Zeno effect. Here an ideal (without
resistance), classical LC oscillating circuit with quick switch
ON-OFF alternation is considered. In the limit of the infinite
series of alternative short electrical current regime (switch in
the ON state) and no-current regime (current breaking by quick
switch ON-OFF state alternation) given LC circuit will never
oscillate. Obviously, it represents a classical electro-dynamical
Zeno effect deeply analogous to quantum Zeno effect. All this
admits a general definition of the Zeno effect that includes both
quantum and classical cases (without any classical interpretation
of the quantum Zeno effect or quantum interpretation of the
classical Zeno effect).
\end {abstract}

In this work a simple classical analog of the quantum Zeno effect
will be suggested. Namely, as it is well known, in the quantum
mechanics, in the limit of the infinite series of alternative
short dynamical evolution and measurement, an unstable quantum
system will never decay, that is called quantum Zeno effect. It is
usual opinion that Zeno effect can exist within quantum mechanics
only. Here an ideal (without resistance), classical LC oscillating
circuit with quick switch ON-OFF alternation will be considered.
In the limit of the infinite series of alternative short
electrical current regime (switch in the ON state) and no-current
regime (current breaking by quick switch ON-OFF state alternation)
given LC circuit will never oscillate. Obviously, it represents a
classical electro-dynamical Zeno effect deeply analogous to
quantum Zeno effect. All this admits a general definition of the
Zeno effect that includes both quantum and classical cases
(without any classical interpretation of the quantum Zeno effect
or quantum interpretation of the classical Zeno effect). More
precisely, Zeno effect can be analogously defined by the infinite
series of the two alternative processes in the classical as well
as in the quantum physics. But, within classical physics
alternative processes, current regime and no-current regime,
necessary represent simple and complex dynamical evolution. On the
other hand, within quantum mechanics second process, i.e.
measurement, cannot at all be considered as the generalized first
process, i.e. generalized dynamical evolution.

Misra and Sudarshan formulated theoretically quantum Zeno paradox
[1], [2] in the following way.

Namely, according to standard quantum mechanical formalism [5],[2]
there are two principally different ways of the changing of the
state of a quantum system. First one represents the unitary (that
conserves superposition), deterministic quantum mechanical
dynamical evolution (presented by Schr$\ddot{o}$dinger equation).
It can be realized during arbitrary small or large time interval
for the quantum system without measurement process. Second one
represents (postulated by von Neumann) the collapse, i.e.
probabilistic superposition breaking by measurement. Collapse,
i.e. measurement realization needs some finite time interval
(determined by Heisenberg uncertainty relations). But formally,
without diminishing of the generality of basic conclusions, it can
be considered that collapse, i.e. measurement appears
instantaneously.

Misra and Sudarshan considered an unstable quantum system, with
total Hamiltonian $\hat {H}$, in the initial, non-decayed quantum
state $|N>$. Initial state quantum mechanically dynamically
(according to Schr$\ddot{o}$dinger equation) evolves, during some
short time interval $[0,t]$, in the final state $|F> = \exp[\frac
{\hat {H}t}{i\hbar}] |N>$ representing a superposition of the
non-decayed, $|N>$, and decayed quantum state, $|D>$. (Roughly
speaking it can be said that final superposition represents the
oscillating process between decayed and non-decayed quantum
state.) This final state, for given short time interval, can be
approximated by its second order Taylor expansion
\begin {equation}
      |F> \simeq (1 - \frac {\hat{H}^{2}t^{2}}{2\hbar^{2}} - \frac {i\hat{H}t}{\hbar})|N>.
\end {equation}
Measurement of the decay, realized in time moment $t$, on the
quantum system in the final state, can detect the non-decayed
state with quantum mechanical probability
\begin {equation}
  w_{N}(t) = (1 -\frac {\Delta \hat{H}^{2}t^{2}}{\hbar^{2}})
\end {equation}
where $\Delta \hat{H}^{2}= <N| \hat{H}^{2}|N>-<N| \hat{H}|N>^{2}$.
Obviously, (2) does not hold linear terms (proportional to $t$).
For this reason (1) represents the minimal non-trivial
approximation of $|F>$.

Suppose now that small time interval $[0,t]$ is divided in the $n$
equivalent subintervals with the same length $\frac {t}{n}$, where
$n$ represents some natural number.  Suppose too that on the end
of any of given time subintervals decay measurement is realized.
Since any measurement is (formally) instantaneous whole interval
with length $t = n \frac {t}{n}$ refers on the dynamical
evolutions only. Then probability that quantum system will be
non-decayed after whole time interval, i.e. in the time moment $t$
equals
\begin {equation}
     w_{Nn} (t) = (1 -\frac {\Delta \hat{H}^{2}(\frac {t}{n})^{2}}{\hbar^{2}})^{n} = (1 - (\frac {t}{n\tau})^{2})^{n}\simeq 1 - (\frac {t}{\tau})^{2}(\frac {1}{n})
\end {equation}
where $\tau = \frac {\hbar}{\Delta \hat {H}}$ represents
characteristic time parameter. Obviously, given probability, in
the limit when $n$ tends toward infinity, tends toward 1, which
means that unstable quantum system will not decay at all.
Metaphorically speaking a watched pot never boils.

Previous statement represents seemingly a paradoxical conclusion.
Namely, as it has been discussed, during whole time interval
$[0,t]$, moment by moment, non-trivial quantum mechanical
dynamical evolution occurs. However, at the end of whole time
interval effectively there is none dynamical effects. But,
strictly speaking, there is none real paradox. All theoretical
predictions are realized strictly according to standard quantum
mechanical formalism on the one hand. On the other hand they are
in the excellent agreement with experimental results [3], [4].
Paradox exists only within naïve intuitive suppositions that
measurement (i.e. interaction of the system with measuring
apparatus) has the form of quantum mechanical dynamical evolution.

Consider now a well-known classical LC oscillating circuit with
conductive tube with inductivity $L$ and condenser with capacity
$C$. Also, suppose that initial electrical charge on the condenser
equals $q_{0}$. Electro-dynamics of this circuit is given by
second Krichhoff rule (including self-inductivity term), i.e.
equation $ - L\frac {d^{2}q}{dt^{2} }= \frac {1}{C}q$ with simple
solution $q = q_{0}\ cos [\omega t]$ where $\omega = (LC)^{-\frac
{1}{2}}$. For sufficiently small t given solution can be
approximated by
\begin {equation}
       q \simeq q_{0} (1 - \frac {\omega^{2}t^{2}}{2}) = q_{0} (1 - \frac {t^{2}}{\tau^{2}}).
\end {equation}
where $\tau = 2^{-\frac {1}{2}}\cdot \omega$ represents
characteristic time parameter.

Suppose that in given LC circuit is a switch. It can be very
quickly, practically instantaneously, settled in the state ON or
in the state OFF. In the first case there is electrical charge
change, i.e. electrical current. In the second case there is no
electrical charge change, i.e. electrical current. Suppose that
switch has bee initially in the ON state i.e. in the current
regime. Suppose too that in the time moment $t$ switch turns out
in the OFF state, i.e. in the no-current or broken current regime.
It implies that in the final moment electrical charge is described
by (4).

Suppose now that small time interval $[0,t]$ is divided in the $n$
equivalent subintervals with the same length $\frac {t}{n}$, where
n represents some natural number.  Suppose again that switch has
bee initially in the ON state. Suppose too that on the end of any
of given time subintervals switch be instantaneously settled
firstly in the OFF state and secondly again in the ON state. Since
any measurement is (formally) instantaneous whole interval with
length $t = n \frac {t}{n}$ refers on the dynamical evolutions
only. Then electrical charge at the end of the whole time
interval, i.e. in time moment t equals
\begin {equation}
     q_{n}(t) = q_{0}(1 - (\frac {t}{n\tau})^{2}) ^{n} \simeq q_{0}(1 - (\frac {t}{\tau})^{2}(\frac
     {1}{n}))
\end {equation}
Obviously, given electrical charge, in the limit when $n$ tends
toward infinity, tends toward 1, which means that electrical
current will not flow at all. Metaphorically speaking finger in
the toaster will never get burned.

Previous statement represents seemingly a paradoxical conclusion.
Namely, as it has been discussed, during whole time interval
$[0,t]$, moment by moment, non-trivial classical electro dynamical
evolution occurs. However, at the end of whole time interval
effectively there is none electro dynamical effects. But, strictly
speaking, there is none real paradox. All theoretical predictions
are realized strictly according to standard classical
electro-dynamics on the one hand. On the other hand they are in
the excellent agreement with experimental results. Paradox exists
only within naïve intuitive suppositions that interaction between
switch and LC oscillating circuit has form of the LC (without
switch or with switch constantly in the ON state) electro
dynamical evolution.

On the basis of the well-known analogy between classical
electro-dynamical LC oscillating circuit and classical mechanical
linear harmonic oscillator, LHO, it can be concluded that on the
LHO with corresponding quick breaking (stopping) mechanism a
classical mechanical Zeno effect can appear too (which will not be
analyzed explicitly).

All this admits the general definition of the Zeno effect that
includes both quantum and classical cases.

Zeno effect occurs on a complex physical system with two
sub-systems, when the following three conditions are satisfied.

Z1:  Dynamical evolution (first process) on the isolated first
sub-system, in the lowest non-trivial approximation for a small
time interval $[0,t]$, is proportional to $(1 - (\frac
{t}{\tau})^{2}) $ where $\tau$ represents characteristic time
parameter.

Z2:  Quick, or, formally, instantaneous, interaction (second
process) of the first with second sub-system stops dynamical
evolution (first process) on the isolated first sub-system.

Z3:  In the mentioned small time interval $[0,t]$ there is a
tending toward infinity series of the alternative first and second
process.

Thus, Zeno effect can be completely analogously defined by the
infinite series of the two alternative processes in the classical
as well as in the quantum physics. But, within classical physics
alternative processes, first (current regime) and second
{no-current regime), represent simple and complex dynamical
evolution. On the other hand, within quantum mechanics second
process, i.e. measurement, cannot at all be considered as the
generalized first process, i.e. generalized dynamical evolution.
Namely, according to remarkable Bell theorem [6], and
corresponding experimental data [7], presentation of the
measurement by any generalized dynamical evolution leads necessary
toward implausible superluminal effects. There is only one
physical possibility without implausible superluminal effects,
that measurement be presented as a form of the non-dynamical phase
transition (with spontaneous superposition breaking) [8], [9].

\vspace{1.5cm}

{\Large \bf References}

\begin {itemize}

\item [[1]] B. Misra, C. J. G. Sudarshan, J. Math. Phys. {\bf 18} (1977) 756
\item [[2]] D. J. Griffiths, {\it Introduction to Quantum Mechanics} (Prentice Hall, Inc., New Jersey, 1995)
\item [[3]] W. M. Itano, D. J. Heinsen, J. J. Bokkinger, D. J. Wineland, Phys. Rev. A {\bf 41} (1999) 2995
\item [[4]] M.C. Fischer, Gutierrez-Medina, M. G. Raizen, Phys. Rev. Lett. {\bf 87} (2001)  040402
\item [[5]] J. von Neumann, {\it Mathematische Grundlagen der Quanten Mechanik} (Springer Verlag, Berlin, 1932)
\item [[6]] J. S. Bell, Physics, {\bf 1} (1964) 195
\item [[7]] A. Aspect, P. Grangier, G. Roger, Phys.Rev.Lett. {\bf  47} (1981) 460
\item [[8]] V. Pankovi\'c, M. Predojevi\'c, M. Krmar, {\it Quantum Superposition of a Mirror and Relative Decoherence (as Spontaneous Superposition Breaking)}, quant-ph/0312015
\item [[9]]   V. Pankovi\'c, T. H$\ddot{u}$bsch, M. Predojevi\'c, M. Krmar, {\it From Quantum to Classical Dynamics: A Landau Phase Transition with Spontaneous Superposition Breaking} quant-ph/0409010

\end {itemize}

\end {document}